\documentstyle[amssymb,preprint,aps]{revtex}

\begin{document}
\title{Probing Spin States of Coupled Quantum dots \\
by dc Josephson Current}
\author{{Yu Zhu}$^1${, Qing-feng Sun}$^2$,{\ and }Tsung-han Lin$^{1,*}$}
\address{{\it State Key Laboratory for Mesoscopic Physics and Department of Physics,}%
\\
Peking University,{\small \ }{\it Beijing, 100871, China}$^1$\\
{\it Center for the Physics of Materials and Department of Physics, McGill}\\
University, Montreal, PQ, Canada H3A 2T8$^2$}
\date{}
\maketitle

\begin{abstract}
We propose an idea for probing spin states of two coupled quantum dots
(CQD), by the dc Josephson current flowing through them. This theory
requires weak coupling between CQD\ and electrodes, but allows arbitrary
inter-dot tunnel coupling, intra- and inter- dot Coulomb interactions. We
find that the Coulomb blockade peaks exhibit a non-monotonous dependence on
the Zeeman splitting of CQD, which can be understood in terms of the Andreev
bound states. More importantly, the supercurrent in the Coulomb blockade
valleys may provide the information of the spin states of CQD: for CQD with
total electron number N=1,3 (odd), the supercurrent will reverse its sign if
CQD becomes a magnetic molecule; for CQD\ with N=2 (even), the supercurrent
will decrease sharply around the transition between the spin singlet and
triplet ground states of CQD.
\end{abstract}


PACS numbers: 74.50.+r, 73.40.Gk, 73.20.Dx, 72.15.Nj.

\baselineskip 20pt 

\newpage

Coupled quantum dots (CQD) fabricated in semiconductors, or so called
``artificial molecule'', is an ideal quantum system for the study of
coherent transport. Each quantum dot (QD) contains discrete energy levels
(quantized due to spacial confinement), which can be tuned independently by
the gate voltages. The coupling between the dots is also adjustable,
resulting in either ``ionic molecule'' for weak coupling or ``covalent
molecule'' for strong coupling. The electron-electron interactions play an
essential role in the transport properties of CQD, due to that the electrons
are added to the molecule one by one, and the conductance exhibits Coulomb
blockade oscillations. Such appealing system has attracted a lot of
experimental and theoretical attentions,\ such as the observations of
Coulomb blockade in double or triple quantum dots \cite{CB1,CB2,CB3,CB4};
the measurement of microwave spectroscopy of a quantum dot molecule and the
comparison with the time dependent theory \cite{MW1,MW2,MW3}; the analysis
of the additional spectra of CQD\ in the magnetic field by using single
electron capacitance spectroscopy \cite{local1}; the investigations of
electronic correlations and Kondo effect in CQD\ by two impurity Anderson
model \cite{Kondo1,Kondo2}, and much more.

Nevertheless, the spin -dependent transport through CQD are less addressed 
\cite{spin1,spin2}. For a single QD, it was proposed recently that QD in the
Coulomb blockade regime may act as an efficient spin filter in the presence
of magnetic field \cite{spinfilter}. For coupled double QDs, the spin
related physics is even more rich. Consider CQD filled with two excess
electrons, each dot occupied by one of them because of the large on-site
Coulomb repulsion. CQD\ in this case may have four spin configurations: $%
\left| 1\uparrow 2\uparrow \right\rangle $, $\left| 1\downarrow 2\downarrow
\right\rangle $, $\left| 1\uparrow 2\downarrow \right\rangle $, $\left|
1\downarrow 2\uparrow \right\rangle $, as shown in the lower part of Fig.1.
Due to the inter-dot tunneling processes, $\left| 1\uparrow 2\downarrow
\right\rangle $ and $\left| 1\downarrow 2\uparrow \right\rangle $ form a
spin singlet state with lower energy. In the presence of magnetic field,
however, the Zeeman energy may overwhelm the singlet binding energy, and $%
\left| 1\uparrow 2\uparrow \right\rangle $ or $\left| 1\downarrow
2\downarrow \right\rangle $ becomes energetically favorable. Thus, CQD may
have either a magnetic or a non-magnetic ground state, depending on the
external magnetic field. The remaining problem is how to detect these states
of CQD. We propose that by attaching CQD\ to two superconductors (S), spin
states of CQD\ can be probed by measuring the dc Josephson current. The idea
is illustrated in the upper part of Fig.1, and hereafter the system is
referred to as S-CQD-S. It is well known that the dc Josephson current is
sensitive to the spin polarization of the weak link area of the junction.
For the Josephson junction coupled by an Anderson impurity (or S-QD-S
heterostructure), theories show that the critical current will reverse its
sign if the impurity (or QD) is singly occupied and therefore becomes a
magnetic dot \cite{sds1,sds2,sds3}. Recent experiment also reported the
observation of the $\pi $-junction transition, in a Josephson junction
consisting of superconducting banks and a weakly ferromagnetic interlayer 
\cite{Pi-J}.

Motivated by the above facts and ideas, we will investigate the supercurrent
flowing through S-CQD-S system. Different from some of the previous works,
e.g. \cite{inf-U}, we take $U\thicksim \Delta $ rather than $U\gg \Delta $,
so that the Andreev reflection (AR) process \cite{Andreev,btk} and the
Coulomb blockade (CB) effect may ``combine'' with each other ($U$ is the
constant of the intra-dot Coulomb interaction, $\Delta $ is the gap of the
superconducting electrodes, and the choice of $U\thicksim \Delta $ is
physically reasonable). In the calculation, we take into account not only
intra-dot but also inter-dot Coulomb interactions, both of which are found
to be important in the fitting of the experimental data \cite{CB4}.

The suggested S-CQD-S system can be described by the following Hamiltonian,

\begin{equation}
H=H_L+H_{DD}+H_R+H_T\;\;,
\end{equation}
in which 
\begin{eqnarray}
H_{DD} &=&\sum_\sigma E_{1\sigma }n_{1\sigma }+U_1n_{1\uparrow
}n_{1\downarrow }+\sum_\sigma E_{2\sigma }n_{2\sigma }+U_2n_{2\uparrow
}n_{2\downarrow }+ \\
&&U_{12}(n_{1\uparrow }+n_{1\downarrow })(n_{2\uparrow }+n_{2\downarrow
})+t\sum_\sigma (c_{1\sigma }^{\dagger }c_{2\sigma }+c_{2\sigma }^{\dagger
}c_{1\sigma })\;\;,  \nonumber
\end{eqnarray}
is for the CQD, modelled by two impurities coupled by inter-dot interaction
and inter-dot tunneling ($n_{i\sigma }\equiv c_{i\sigma }^{\dagger
}c_{i\sigma }$ is the particle number operator); $H_L$ and $H_R$ are
standard BCS Hamiltonians for the s-wave superconducting electrodes, with
superconducting phases $\phi _L=\phi /2$ and $\phi _R=-\phi /2$,
respectively \cite{remark1}; and $H_T$ is tunneling Hamiltonian , connecting
the three parts together.

We begin with the study of the eigen states of isolated CQD. Notice that $%
H_{DD}$ can be exactly diagonalized in the particle number representation,
i.e., in the set of 16 bases $\left| 0\right\rangle ,$ $\left| 1\uparrow
\right\rangle ,\cdots \;\left| 1\uparrow 1\downarrow 2\uparrow 2\downarrow
\right\rangle $ \cite{diagonal}. Due to the conservation of particle number
and the conservation of spin in $H_{DD}$, the 16$\times $16 space can be
divided into several sub-spaces, 
\begin{equation}
16=(1)_{\text{N=0}}+(2+2)_{\text{N=1}}+(1+1+4)_{\text{N=2}}+(2+2)_{\text{N=3}%
}+(1)_{\text{N=4}}\;\;.
\end{equation}
We have special interest in the N=2 sub-space, in which $H_{DD}$ is
expressed as 
\begin{equation}
\begin{tabular}{|c|c|c|c|c|c|}
\hline
$\tilde{E}_1\;\;$ &  &  &  &  &  \\ \hline
& $\tilde{E}_2\;\;$ &  &  &  &  \\ \hline
&  & $\tilde{E}_3\;\;$ & 0 & t & t \\ \hline
&  & 0 & $\tilde{E}_4\;\;$ & t & t \\ \hline
&  & t & t & $\tilde{E}_5\;\;$ & 0 \\ \hline
&  & t & t & 0 & $\tilde{E}_6\;\;$ \\ \hline
\end{tabular}
\end{equation}
with $\tilde{E}_1=E_{1\uparrow }+E_{2\uparrow }+U_{12}$, $\tilde{E}%
_2=E_{1\downarrow }+E_{2\downarrow }+U_{12}$, $\tilde{E}_3=E_{1\uparrow
}+E_{2\downarrow }+U_{12}$, $\tilde{E}_4=E_{2\uparrow }+E_{1\downarrow
}+U_{12}$, $\tilde{E}_5=E_{1\uparrow }+E_{1\downarrow }+U_1$, and $\tilde{E}%
_6=E_{2\uparrow }+E_{2\downarrow }+U_2$. For the case of identical dots \cite
{remark2}, i.e., $U_1=U_2\equiv U$, $U_{12}\equiv V$, $E_{1\uparrow
}=E_{2\uparrow }\equiv E_0-h$, $E_{1\downarrow }=E_{2\downarrow }\equiv
E_0+h $, the eigen solution of $H_{DD}$ has a simple form: \{$\left|
1\uparrow 2\uparrow \right\rangle $, $2E_0-2h+V$\}, \{$\left| 1\downarrow
2\downarrow \right\rangle $, $2E_0+2h+V$\}, \{$\left| \alpha
_{+}\right\rangle $, $2E_0+V $\}, \{$\left| \beta _{+}\right\rangle $, $%
2E_0+U$\}, \{$\left| a_{-}\right\rangle $, $2E_0+V-E_b$\}, \{$\left|
b_{-}\right\rangle $, $2E_0+U+E_b$\}, in which $\left| \alpha _{\pm
}\right\rangle \equiv \frac 1{\sqrt{2}}\left( \left| 1\uparrow 2\downarrow
\right\rangle \mp \left| 1\downarrow 2\uparrow \right\rangle \right) $, $%
\left| \beta _{\pm }\right\rangle \equiv \frac 1{\sqrt{2}}\left( \left|
1\uparrow 1\downarrow \right\rangle \mp \left| 2\uparrow 2\downarrow
\right\rangle \right) $, $\left| a_{-}\right\rangle $ and $\left|
b_{-}\right\rangle $ are the two new states perturbed from $\left| \alpha
_{-}\right\rangle $ and $\left| \beta _{-}\right\rangle $. $\left| 1\uparrow
2\uparrow \right\rangle $, $\left| 1\downarrow 2\downarrow \right\rangle $
and $\left| \alpha _{+}\right\rangle $ are corresponding to the S=1 triplet
states with m=1,-1,0; while $\left| \tilde{\alpha}_{-}\right\rangle $ is the
S=0 singlet state, with a binding energy 
\begin{equation}
E_b=\frac 12\left[ \sqrt{(U-V)^2+(4t)^2}-(U-V)\right] \;\;.
\end{equation}
With the 16 eigenstates of $H_{DD}$, the occupation number of CQD can be
evaluated by using the relation $\left\langle o\right\rangle =Tr(\rho o)$,
in which $\rho =\frac 1Ze^{-\beta H}$ is the density matrix operator. The
left and right insets of Fig.2 shows the occupation number per spin $%
\left\langle n_\sigma \right\rangle \equiv \left\langle n_{1\sigma
}\right\rangle +\left\langle n_{2\sigma }\right\rangle $ vs the resonant
level $E_0$ for different Zeeman splitting $h$. (In practice, $E_0$ can be
tuned by the gate voltage, and $h$ induced by applying an in-plane magnetic
field.) The left inset is for the case of $h<h_c$, while the right is for $%
h>h_c$, where $2h_c\equiv E_b$ depicting the competition between the Zeeman
energy and the singlet binding energy. These curves show that CQD with total
electron number N=1, 3 is easily magnetized in a magnetic field, while CQD
with N=2 favors a spin singlet state and only transfers to a magnetic state
upon a critical magnetic field.

Next, we turn on the weak coupling between S electrodes and CQD. ``Weak
coupling'' means that the supercurrent flowing through CQD only serves as a
probe to provide the information of CQD, without disturbing the quantum
states there. To include the tunneling between the two dots and the physics
of AR, a 4$\times $4 representation is introduced, 
\begin{equation}
{\bf G}^{r,a,<}(\omega )\equiv \langle \langle \left( 
\begin{array}{l}
c_{1\uparrow }| \\ 
c_{1\downarrow }^{\dagger }| \\ 
c_{2\uparrow }| \\ 
c_{2\downarrow }^{\dagger }|
\end{array}
\right) \left( 
\begin{array}{llll}
c_{1\uparrow }^{\dagger } & c_{1\downarrow } & c_{2\uparrow }^{\dagger } & 
c_{2\downarrow }
\end{array}
\right) \rangle \rangle _\omega ^{r,a,<}\;\;.
\end{equation}
Notice that the retarded Green function of the isolated CQD (${\bf g}^r$)
can be constructed exactly by the Lehmann spectral representation, 
\begin{equation}
\langle \langle A|B\rangle \rangle _\omega ^r=\frac 1Z\sum_{nm}\frac{%
e^{-\beta E_n}+e^{-\beta E_m}}{\omega -(E_n-E_m)+\text{i}0^{+}}\left\langle
m|A|n\right\rangle \left\langle n|B|m\right\rangle \;\;,
\end{equation}
in which $n$ or $m$ runs over the 16 eigenstates of $H_{DD}$, and $A$ or $B$
denotes $c_{i\sigma }$ or$\;c_{i\sigma }^{\dagger }$. We assume that the
full Green function of S-CQD-S (${\bf G}^r$) can be derived by the following
approximation \cite{remark3} 
\begin{equation}
{\bf G}^r{\bf =g}^r{\bf +g}^r{\bf \Sigma }^r{\bf G}^r\;\;,
\end{equation}
in which ${\bf \Sigma }^r$ is the self-energy caused by the coupling between
S electrodes and CQD. ${\bf \Sigma }^r$ is obtained as 
\begin{equation}
{\bf \Sigma }^r=\left( 
\begin{array}{cc}
{\bf \Sigma }_L^r & 0 \\ 
0 & {\bf \Sigma }_R^r
\end{array}
\right) \;\;,
\end{equation}
in which 
\begin{eqnarray}
{\bf \Sigma }_\beta ^r &=&-\frac{\text{i}}2\Gamma _\beta \rho (\omega
)\left( 
\begin{array}{cc}
1 & -\frac \Delta {\omega +\text{i}0^{+}}e^{-\text{i}\phi _\beta } \\ 
-\frac \Delta {\omega +\text{i}0^{+}}e^{\text{i}\phi _\beta } & 1
\end{array}
\right) \;\;\;\;(\beta =L,\;R)\;\;,\;\; \\
\rho (\omega ) &\equiv &\frac{\omega +\text{i}0^{+}}{\sqrt{(\omega +\text{i}%
0^{+})^2-\Delta ^2}}\;\;\;\;\;\;\;\;(%
\mathop{\rm Im}%
\sqrt{x}>0)\;\;,\;
\end{eqnarray}
and $\Gamma _{L/R}$ is the coupling strength between left / right electrode
and CQD. Then the dc Josephson current flowing through S-CQD-S is obtained
by the Green function technique as 
\begin{equation}
I=\frac{2e}\hbar \sin \phi \int \frac{d\omega }{2\pi }f(\omega )j(\omega
)\;\;,
\end{equation}
in which 
\begin{equation}
j(\omega )=%
\mathop{\rm Im}%
J(\omega )=%
\mathop{\rm Im}%
\left[ -\frac{\Gamma _L\Gamma _R\Delta ^2}{(\omega +\text{i}0^{+})^2-\Delta
^2}\cdot \frac{({\bf g}^r)_{13}^{-1}({\bf g}^r)_{24}^{-1}}{\det [{\bf (g}^r%
{\bf )}^{-1}{\bf -\Sigma }^r]}\right] \;\;,
\end{equation}
and $f(\omega )=1/(e^{\beta \omega }+1)$ is the Fermi distribution function. 
$\det [{\bf (g}^r{\bf )}^{-1}{\bf -\Sigma }^r]$ in the denominator has
several real roots (with infinitesimal imaginary part -i$0^{+}$) within the
range of $\left| \omega \right| <\Delta $, corresponding to the Andreev
bound states (ABS) in the S-CQD-S system. To avoid the divergence around
these singularities, we adapt the integral path to a V-shaped contour shown
in the middle inset of Fig.2 \cite{remark4}, as a result 
\begin{equation}
I=\frac{2e}\hbar \sin \phi 
\mathop{\rm Im}%
\int_V\frac{d\omega }{2\pi }f(\omega )J(\omega )\;\;.
\end{equation}

Fig.2 shows the critical current $I_c\equiv I(\phi =\frac \pi 2)$ vs $E_0$
with different $h$. The curve of $h=0$, as expected, has four CB peaks,
among which are the valleys of the total electron number from N=0 to N=4. (
The curve is symmetric to $E_0=-(U/2+2V)$ by virtue of the electron and hole
symmetry.) With the increase of $h$, the 1st and 4th peak are gradually
suppressed \cite{remark5}, while the 2nd and 3rd peak exhibit a
non-monotonous dependence on $h$: decrease first and reverse its sign to a
negative peak, then increase again and diminish at sufficient large $h$. To
understand this anomalous $h$ dependence, we plot the curves of $I_c$ vs $h$
in Fig.3a, for a non-interacting S-CQD-S system by setting $U=V=0$. (In this
case, the approximate Eq.(8) becomes an exact one, and ${\bf g}^r$ can be
evaluated explicitly.) For $h=0$, there are two peaks in the curve separated
by $2t$; for $h>0$, $I_c$ at $E_0=0$ also exhibits a non-monotonous $h$
dependence (see the up-right inset). As we know, the supercurrent flowing
through CQD is conducted mainly by the ABS, which have the property that two
adjacent states carry the supercurrent with opposite signs. The spectrums of 
$j(\omega )$ for $h=0,t,2t$ are shown in the insets of Fig.3a. In each
spectrum, there are four ABS denoted by $1^{\pm }$ and $2^{\pm }$ within the
superconducting gap, and continuous spectrum $c^{\pm }$ outside the gap. For 
$h=0$, $1^{\pm }$ and $2^{\pm }$ are distributed symmetrically to the Fermi
surface $\mu =0$. With the increase of $h$ , these states move down toward $%
\omega =-\Delta $. The case of $h=t$ corresponds to three of ABS $1^{\pm }$
and $2^{-}$ are below the Fermi surface, while $h=2t$ corresponds to all of
them below the Fermi surface. By taking account of the Fermi distribution
and the continuous spectrum contribution from $c^{\pm }$, the anomalous $h$
dependence is readily understood. Considering the Coulomb interactions will
induce more ABS in $j(\omega )$, but the $h$ dependence of the peaks in the $%
I_c$ vs $E_0$ curve can be explained in the similar way.

We are more interested in the supercurrent flowing in the CB\ valleys, since
the spin states of CQD\ are well defined there. The curves of $I_c$ vs $h$
and corresponding $m\equiv \left\langle n_{\uparrow }\right\rangle
-\left\langle n_{\downarrow }\right\rangle $ vs $h$ are shown in Fig.3b,
with $E_0$ chosen in the CB valleys of N=1, 2, 3. The two curves marked with
A are typical for the supercurrent flowing in the valley with odd number,
where CQD\ has a net spin. Due to the strong Coulomb interaction, small
magnetic field ($h\thicksim k_BT$) will lead to the transition of CQD from a
non-magnetic molecule to a magnetic one; meanwhile, $I_c$ reverse its sign
and experiences a ``$\pi $-junction'' transition. The two curves marked with
B, in contrast, are typical for the supercurrent flowing in the valley with
even number, where CQD\ may choose a non-magnetic singlet or a magnetic
triplet as its ground state. The transition between them occurs at the
critical value of Zeeman splitting $2h_c=E_b$, with $I_c>0$ for $h<h_c$ and $%
I_c\rightarrow 0^{+}$ for $h>h_c$. Out of our expectation, for $h\gg h_c$, $%
I_c\rightarrow 0^{+}$ instead of $I_c\rightarrow 0^{-}$. Roughly, this might
be interpreted as the addition of two $\pi $-junction transitions, namely,
the supercurrent reverse its sign after flowing through the first magnetic
dot, but returns back after flowing through the second one.

To sum up, we have investigated the dc Josephson current flowing through
S-CQD-S hybrid system. Our calculation is based on the exact diagonalization
of coupled two impurity model and the construction of Green function in the
spectral representation. To avoid the difficulty of solving the exact ABS,
we choose a V-shaped integral contour in the complex $\omega $ plane. In the
numerical study, we find that the dc Josephson current flowing through
S-CQD-S can provide rich information of the spin polarization of CQD. The CB
peaks exhibit a non-monotonous dependence on the Zeeman splitting of CQD,
the current in the CB valleys can be used to probe the spin states of CQD.
We believe that the proposed S-CQD-S system is within the scope of the
update nano-technology in S/2DEG heterostructure, and we are looking forward
to seeing the relevant experiments.

This project was supported by NSFC\ under Grant No. 10074001. 

\smallskip $^{*}$ To whom correspondence should be addressed.


\section*{Figure Captions}

\begin{itemize}
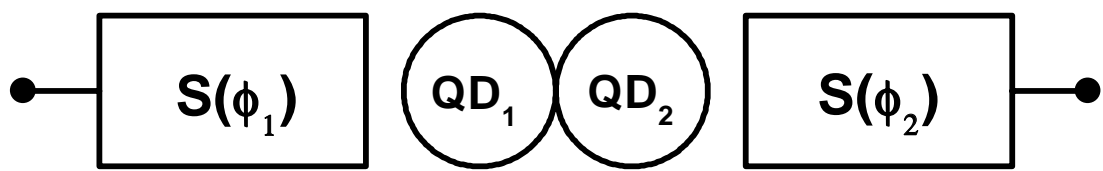
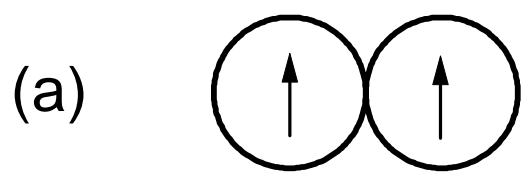
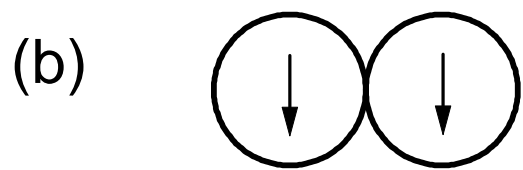
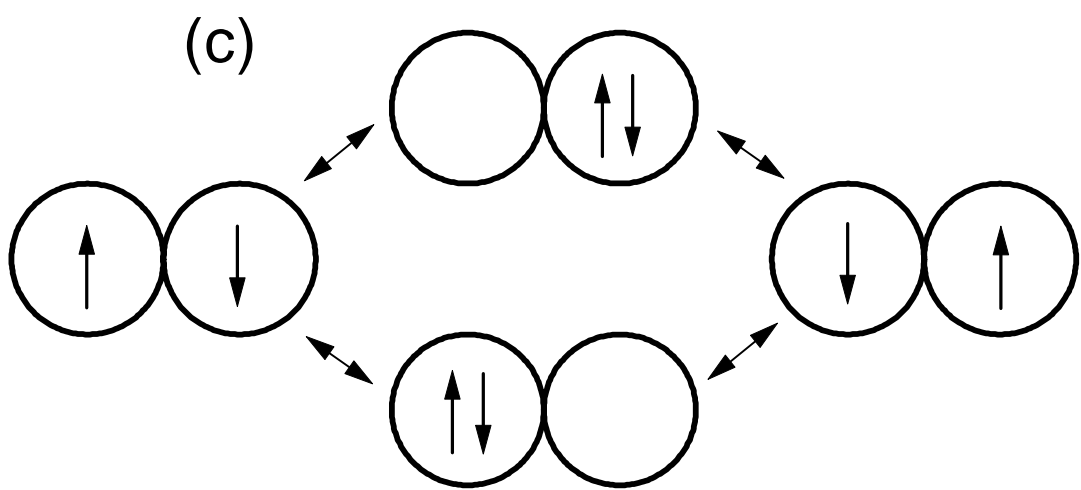
\item[{\bf Fig. 1}]  Upper part: Schematic diagram of the proposed S-CQD-S\
system. Lower part: The four spin configurations of CQD with the total
electron number N=2. (a) and (b) are for $\left| 1\uparrow 2\uparrow
\right\rangle $ and $\left| 1\downarrow 2\downarrow \right\rangle $,
respectively; (c) illustrates that $\left| 1\uparrow 2\downarrow
\right\rangle $ and $\left| 1\downarrow 2\uparrow \right\rangle $ are
coupled via high energy virtual states and forms a spin singlet.

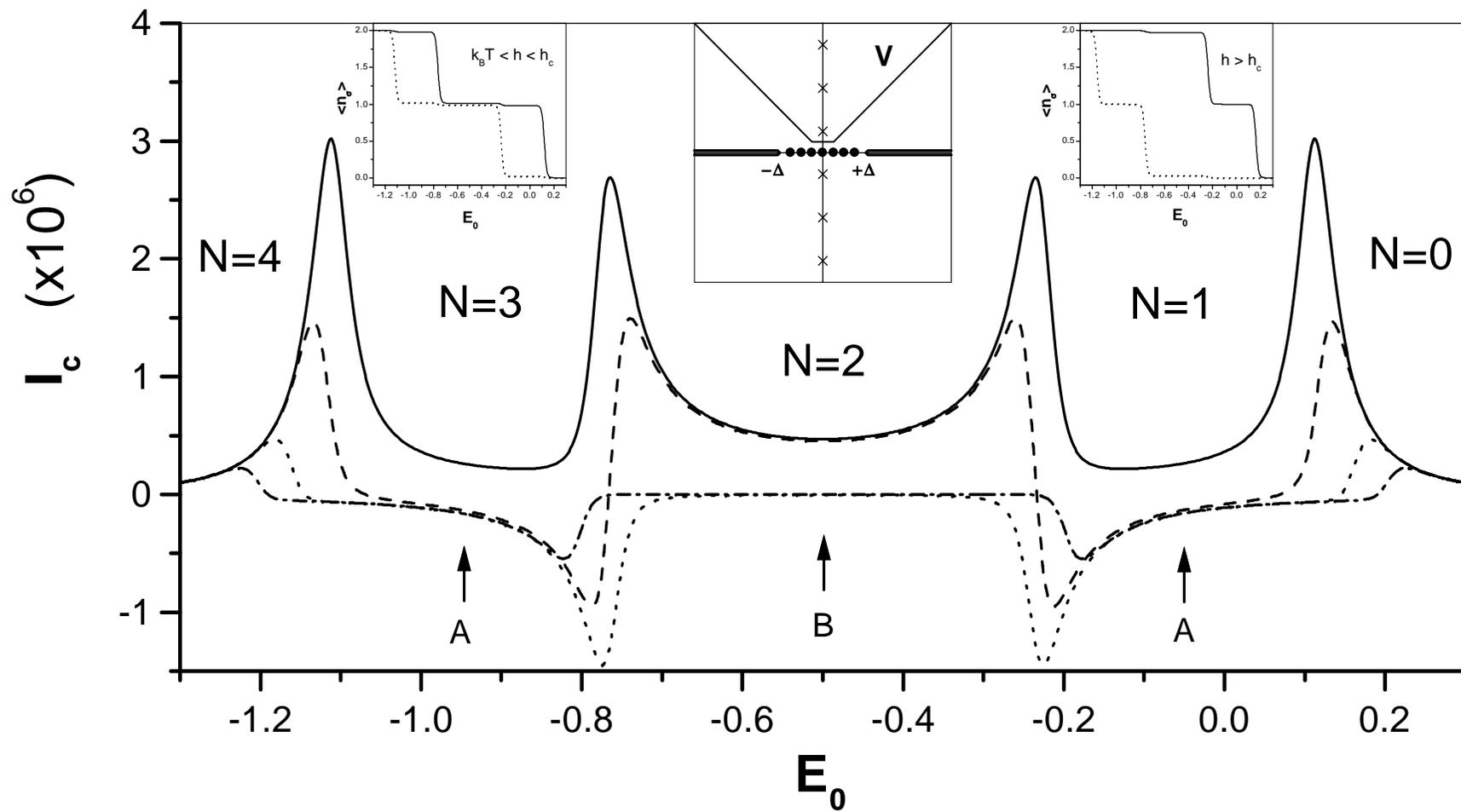
\item[{\bf Fig. 2}]  The critical current $I_c\equiv I(\frac \pi 2)$ vs the
resonant level $E_0$ with different Zeeman splitting $h$, in units of $%
e=\hbar =\Delta =1$. The parameters of CQD are: $U=0.6$, $V=0.2$, $t=0.1$,
due to which $h_c\thickapprox 0.04$. The temperature and coupling strength
are chosen as $k_BT=0.01\gg \Gamma =0.001$, as required by the weak coupling
limit. The solid, dashed, dotted, and dash-dotted curves correspond to $%
h=0.00,\;0.02,\;0.06,\;0.10$, respectively. The left and right insets show
the occupation number per spin $\left\langle n_\sigma \right\rangle $ vs $%
E_0 $ for $h=0.02$ and $h=0.06$, respectively. The middle inset
schematically shows the singularities of $J(\omega )$ and $f(\omega )$ and
the V-shaped integral contour.

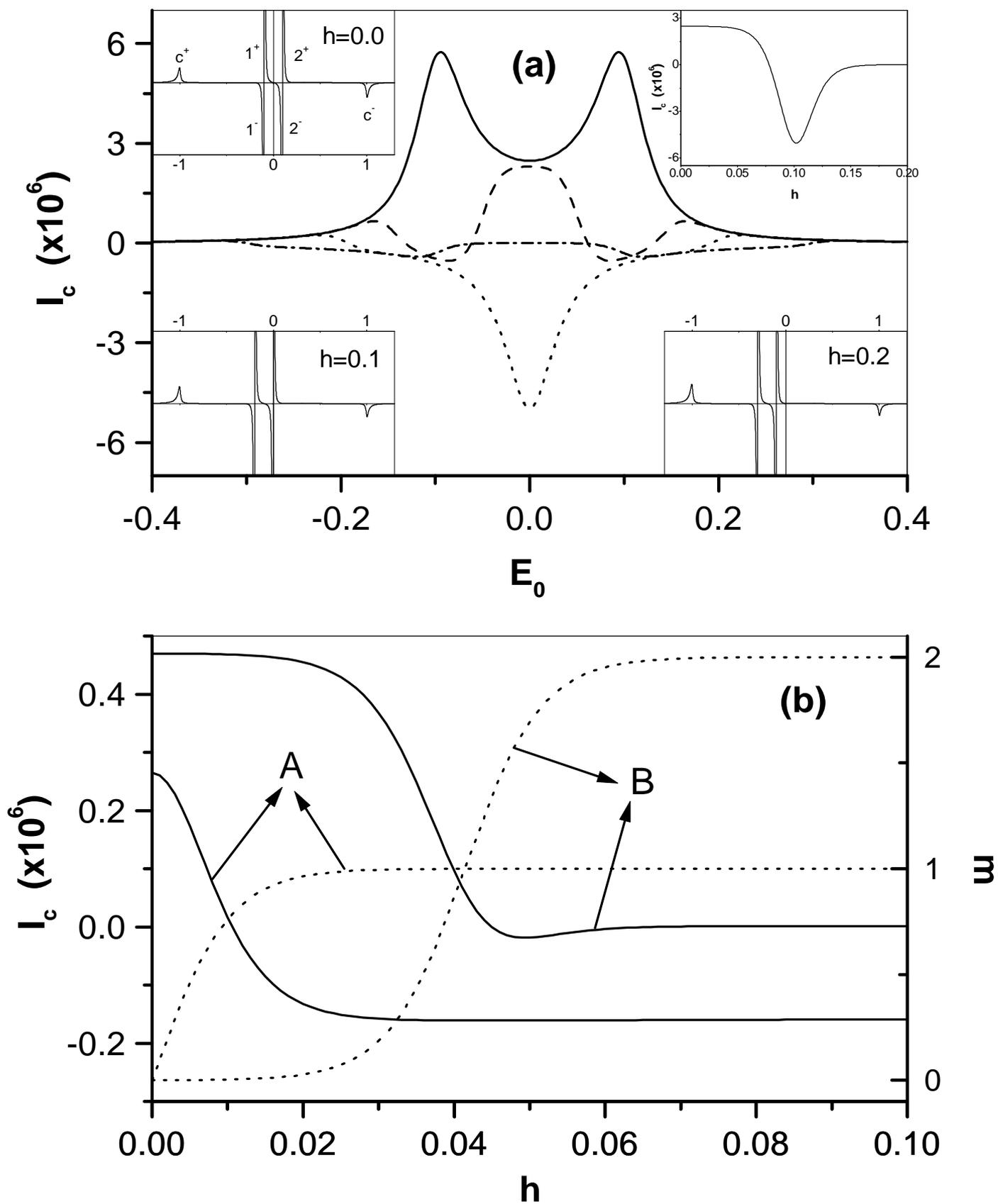
\item[{\bf Fig. 3}]  (a) $I_c$ vs $E_0$ curves with different $h$, for the
case of non-interacting S-CQD-S in which $U=V=0$. The solid, dashed, dotted,
and dash-dotted curves correspond to $h=0.00,\;0.05,\;0.10,\;0.20$,
respectively. Other parameters are the same as those of Fig.2. The up-right
inset shows $I_c$ vs $h$ at $E_0=0$, which exhibits a non-monotonous
dependence. The other three insets show the spectra of $j(\omega )$ at $%
E_0=0 $, with different $h$ marked in the plots. (b) $I_c$ vs $h$ (solid)
and corresponding $m\equiv \left\langle n_{\uparrow }\right\rangle
-\left\langle n_{\downarrow }\right\rangle $ vs $h$ (dotted), in the Coulomb
blockade valleys of N=1, 2, 3. The two curves marked with A are for $%
E_0=-0.95$ (also $E_0=-0.05$), typical for CQD with odd number of electrons;
the other two marked with B are for $E_0=-0.50$, typical for CQD with even
number of electrons.
\end{itemize}


\begin{references}
\bibitem{CB1}  F. R. Waugh $et\;al.$, Phys. Rev. Lett. {\bf 75, }705 (1995).

\bibitem{CB2}  C. Livermore $et\;al.$, Science {\bf 274, }1332 (1996).

\bibitem{CB3}  S. M. Maurer, S. R. Patel, and C. M. Marcus, Phys. Rev. Lett. 
{\bf 83, }1043 (1999).

\bibitem{CB4}  S. D. Lee $et\;al.$, Phys. Rev. B {\bf 62, }R7735 (2000).

\bibitem{MW1}  C. A. Stafford and N. S. Wingreen, Phys. Rev. Lett. {\bf 76, }%
1916 (1996).

\bibitem{MW2}  T. H. Oosterkamp $et\;al.$, Nature (London) {\bf 395, }873
(1998).

\bibitem{MW3}  Q. -f. Sun, J. Wang, and T. -h. Lin, Phys. Rev. B {\bf 61, }%
12643 (2000).

\bibitem{local1}  M. Brodsky $et\;al.$, Phys. Rev. Lett. {\bf 85, }2356
(2000).

\bibitem{Kondo1}  A. Georges and Y. Meir, Phys. Rev. Lett. {\bf 82, }3508
(1999).

\bibitem{Kondo2}  R. Aguado and D. C. Langreth, Phys. Rev. Lett. {\bf 85, }%
1946(2000).

\bibitem{spin1}  C. A. Stafford, R. Kotlyar, and S. D. Sarma, Phys. Rev. B 
{\bf 58, }7091 (1998).

\bibitem{spin2}  Y. Asano, Phys. Rev. B {\bf 58, }1414 (1998).

\bibitem{spinfilter}  P. Recher, E. V. Sukhorukov, and D. Loss, Phys. Rev.
Lett. {\bf 85, }1962 (2000).

\bibitem{sds1}  L. I. Glazman and K. A. Matveev, JETP Lett. {\bf 49, }659
(1989).

\bibitem{sds2}  B. I. Spivak and S. A. Kivelson, Phys. Rev. B {\bf 43, }3740
(1991).

\bibitem{sds3}  S. Ishizaka, J. sone, and T. Ando, Phys. Rev. B {\bf 52, }%
8358 (1995).

\bibitem{Pi-J}  V. V. Ryazanov $et\;al.$, Phys. Rev. Lett. {\bf 86, }2427
(2001).

\bibitem{inf-U}  M. -S. Choi, C. Bruder, and D. Loss, Phys. Rev. B {\bf 62, }%
13569 (2000).

\bibitem{Andreev}  A. F. Andreev, Zh. Eksp. Teor. Fiz. {\bf 46, }1823 (1964)
[Sov. Phys. JETP {\bf 19, }1228 (1964)].

\bibitem{btk}  G. E. Blonder, M. Tinkham, and T. M. Klapwijk, Phys. Rev. B 
{\bf 25, }4515 (1982).

\bibitem{remark1}  The chemical potentials of both superconducting
electrodes are set as zero, i.e., $\mu _L=\mu _R=0$.

\bibitem{diagonal}  G. Chen, G. Klimeck, and S. Datta, Phys. Rev. B {\bf 50, 
}8035 (1994).

\bibitem{remark2}  In all the numerical study, we assume that the two dots
are identical, although our formula is applicable to general cases.

\bibitem{remark3}  The approximation is expected to be equivalent to a
proper truncation of equation of motion, which has been tested in N-CQD-N,
see C. Niu, L. -J. Liu, and T. -H. Lin, Phys. Rev. B {\bf 51, }5130 (1995).

\bibitem{remark4}  Notice that all of the singularities of $J(\omega )$ lie
in the lower half-plane.

\bibitem{remark5}  Since N=0 and N=4 correspond to empty and fully occupied
CQD, they are not typical for a CB valley, and hence the 1st and 4th peak in
Fig.2 are not typical for a CB peak. \newpage
\end{references}
\end{document}